\documentclass[12pt]{article}
\usepackage[dvips]{color}
\usepackage{epsfig}
\usepackage{amsmath}
\usepackage{graphicx}

\begin{document}
\title{\bf Viscous Modified Cosmic Chaplygin Gas Cosmology}
\author{{B. Pourhassan\thanks{Email: b.pourhassan@umz.ac.ir}}\\
{\small {\em Department of Physics, I.H.U., Tehran, Iran}}}
\maketitle
\begin{abstract}
\noindent In this paper we construct modified cosmic Chaplygin gas which has viscosity.
We use exponential function method to solve non-linear equation and obtain time-dependent dark energy density.
Then discuss Hubble expansion parameter and scale factor and fix them by using observational data. We also investigate stability of this
theory.
{\bf Pacs Number:} 95.35.+d; 95.85.-e; 98.80.-k\\
\end{abstract}
\newpage
\tableofcontents
\newpage
\section{Introduction}
It is believed that the most part of Universe filled with dark matter and dark energy. Therefore, dark energy and related topics are important subjects to
study in theoretical physics and cosmology. An important problem is determining nature of dark Universe. It is found that the dark matter may be consists
of neutrinos [1] axions [2] or WIMPs (weak interactive massive particles) [3]. In that case there are several ways to specify the nature of the dark
Universe. For example studying time-dependent density help to give information about dark matter [4] and dark energy [5].\\
After that we need a model to describe dark Universe. In that case there are some phenomenological and theoretical models which based on discovery of the
accelerating expansion of the Universe [6, 7]. Some of the famous phenomenological models for dark energy briefly explained below.\\
The cosmological constant and its generalizations are the simplest way to modeling the dark energy [8]. Another candidates for the dark energy are
scalar-field dark energy models. A quintessence field [9] is a scalar field with standard kinetic term, which minimally coupled to gravity. In that case
the action may has a wrong sign kinetic term (minus instead of plus), which the scalar field is called phantom or ghost [10]. While there is a quantum
instability in the phantom models but this model to be consistent with CMB observations [11]. Combination of the quintessence and the phantom is known as
the quintom, which is another model for dark energy [12].\\
Extension of kinetic term in Lagrangian yields to a more general frame work on field theoretic dark energy, which is called k-essense [13, 14]. A singular
limit of k-essense is another model, named Cuscuton [15]. This model has an infinite propagating speed for linear perturbations, however causality is still
valid.\\
The most general form for a scalar field with second order equation of motion is the Galileon field which could behave as dark energy [16]. Another
extension of these models is called the ghost condensation, which also solved the quantum instability of phantom dark energy [17]. There are also various
studies in holographic dark energy models (see Refs. [18-22]).\\
However, presence of a scalar field is not only requirement of the transition from a Universe filled with matter to an exponentially expanding Universe.
The matter components in cosmology are written in terms of fluids, so most of dark energy models have fluid description. Therefore, Chaplygin gas (CG) used
as an exotic type of fluid, which is a model for dark energy [23, 24]. This model based on Chaplygin equation of state [25] to describe the lifting force
on a wing of an air plane in aerodynamics. The CG was not consistent with observational data of SNIa, BAO, CMB, and so on [26-29]. Therefore, an extension
of CG model proposed [30, 31], which is called generalized Chaplygin gas (GCG), and indeed proposed unification of dark matter and dark energy. However,
observational data ruled out such a proposal. Then, GCG extend to the modified Chaplygin gas (MCG) [32]. There is still more extension such as generalized
cosmic Chaplygin gas (GCCG) [33]. In this paper we deal with next extension which is modified cosmic Chaplygin gas [34].\\
On the other hand bulk viscosity plays an important role in the evolution of the Universe. The idea that Chaplygin gas may has viscosity first proposed by
the Ref. [35] and then developed by [34, 36, 37, 38, 39].\\
In the Ref. [35], for the first time, GCG with bulk viscosity studied. In the Ref. [34] we indeed extend Ref. [35] to the case of viscous MCG for the
special case of $\alpha=1/2$ in equation of state formula and $k=0$ in FRW metric. Then, in the Ref. [37] we extend our previous work to the case of MCCG
and calculate time-dependent dark energy density and discussed stability of system. In the Ref. [38] viscous CG in non-flat FRW Universe has been studied
and time-dependent density of dark energy obtained under assumption of cosmic expansion without acceleration or deceleration which means power law scale
factor. Then in the Ref. [39] we extend our previous work to the case of arbitrary $\alpha$ instead of $\alpha=1/2$ and studied viscous GCG. In all cases
the time-dependent density calculated approximately, and under some assumption for simplicity.\\
Now, in this paper we would like to study viscous MCCG with
arbitrary $\alpha$ and use powerful tools to obtain more exact
solutions. Indeed, in order to solve non-linear equation we use
exponential function method which is more appropriate than previous
approximate methods. Furthermore we investigate evolution of scale
factor and try to compare our results with observational data, and
also discuss stability of system by using speed of sound in fluids.
This paper is organizing as the following.

\section{Review of Chaplygin gas model}
One of the recent cosmological models which is based on the use of exotic type of perfect fluid suggests that our Universe filled with the Chaplygin gas
with the following equation of state [38, 40],
\begin{equation}\label{s1}
p=-\frac{B}{\rho},
\end{equation}
where $B$ is a positive constant. The equation (1) introduced by Chaplygin to describe the lifting force on an airplane wing [25]. Chaplygin gas is also
interesting subject of holography [41], string theory [42], and supersymmetry [43]. It is also possible to study FRW cosmology of a Universe filled with CG
[44]. CG equation of state (1) has been generalized to the form [24, 45, 46, 47],
\begin{equation}\label{s2}
p=-\frac{B}{\rho^{\alpha}},
\end{equation}
with $0<\alpha\leq1$, which is called GCG, and is also interesting from holography point of view [48]. As we can see from the equation (2), the GCG is
corresponding to almost dust ($p=0$) at high density which is not agree completely with our Universe. Therefore, MCG with the following equation of state
introduced [32, 49, 50],
\begin{equation}\label{s3}
p=\gamma\rho-\frac{B}{\rho^{\alpha}},
\end{equation}
where $\gamma$ is a positive constant. This model is more appropriate choice to have constant negative pressure at low energy density and high pressure at
high energy density. The special case of $\gamma=1/3$ is the best fitted value to describe evolution of the Universe from radiation regime to the
$\Lambda$-cold dark matter regime. Also, the MCG with $\gamma=0.085$ and $\alpha=1.724$ is coincident with some observational data [51]. The next extension
performed by the Ref. [33] where the GCCG introduced by the following equation of state,
\begin{equation}\label{s4}
p=-\frac{1}{\rho^{\alpha}}\left[C+(\rho^{1+\alpha}-C)^{-\omega}\right],
\end{equation}
where $C$ is a constant. In the case of $\omega=0$ one can write $C=B-1$. The speciality of this model is stability so the theory is free from unphysical
behaviors even when the vacuum fluid satisfies the phantom energy condition. In this paper we construct the MCCG with the following equation of state,
\begin{equation}\label{s5}
p=\gamma\rho-\frac{1}{\rho^{\alpha}}\left[\frac{B}{1+\omega}-1+(\rho^{1+\alpha}-\frac{B}{1+\omega}+1)^{-\omega}\right].
\end{equation}
Our main goal is using above equation of state to study FRW bulk viscous cosmology, therefore we need to review FRW bulk viscous cosmology and then add
viscosity to MCCG.
\section{FRW bulk viscous cosmology}
As we know the Friedmann-Robertson-Walker (FRW) Universe is described by the following metric,
\begin{equation}\label{s6}
ds^2=-dt^2+a(t)^2(\frac{dr^2}{1-kr^{2}}+r^{2}d\Omega^{2}),
\end{equation}
where $d\Omega^{2}=d\theta^{2}+\sin^{2}\theta d\phi^{2}$, and $a(t)$ represents the scale factor. The $\theta$ and $\phi$ parameters are the usual
azimuthal and polar angles of spherical coordinates, with $0\leq\theta\leq\pi$ and $0\leq\phi<2\pi$. The coordinates ($t, r, \theta, \phi$) are called
co-moving coordinates. The constant $k$ defined curvature of space so, $k=0, 1$, and $-1$ represents flat, closed and open spaces respectively. Our
interest in this paper is the first case namely $k=0$. In that case the Einstein equation is given by,
\begin{equation}\label{s7}
R_{\mu\nu}-\frac{1}{2}g_{\mu\nu}R=T_{\mu\nu},
\end{equation}
where we assumed $c=1$, $8\pi G = 1$ and $\Lambda=0$. It is assumed that our Universe is filled with the MCCG which plays role of dark energy with equation
of state (5). Using the line element (6) and the Einstein equation (7), the energy-momentum tensor corresponding to the bulk viscous fluid is given by the
following relation,
\begin{equation}\label{s8}
T_{\mu\nu}=(\rho+\bar{p})u_{\mu}u_{\nu}-\bar{p}g_{\mu\nu},
\end{equation}
where $\rho$ is the dark energy density and $u^{\mu}$ is the velocity vector with normalization condition $u^{\mu}u_{\nu}=-1$. Also,
\begin{equation}\label{s9}
\bar{p}=p-3\zeta H,
\end{equation}
is the total pressure which involves the proper pressure $p$, given by equation of state (5), bulk viscosity coefficient $\zeta$ and Hubble expansion
parameter $H=\dot{a}/a$. It is also assumed that the dark energy is conserved with the following conservation equation,
\begin{equation}\label{s10}
\dot{\rho}+3H(\bar{p}+\rho)=0.
\end{equation}
In the next section we write field equations of viscous MCCG, and then try to solve them for dark energy density.

\section{Viscous modified cosmic Chaplygin gas}
By using the relations of previous section one can obtain the following field equations,
\begin{equation}\label{s11}
H^{2}=\frac{\rho}{3},
\end{equation}
and
\begin{equation}\label{s12}
\dot{H}+H^{2}=-\frac{\rho}{6}-\frac{\bar{p}}{2},
\end{equation}
where dot denotes derivative with respect to cosmic time $t$. The energy-momentum conservation law (10) reduced to the following equation,
\begin{equation}\label{s13}
\dot{\rho}+3(\rho+p)H-9\zeta H^{2}=0,
\end{equation}
where $\partial_{\nu}T^{\mu\nu}=0$ is used. The equation (11) can be easily extended to the case of non-flat Universe and non-vanishing cosmological
constant,
\begin{equation}\label{s14}
H^{2}=\frac{\rho}{3}-\frac{k}{a^{2}}+\frac{\Lambda}{3}.
\end{equation}
As we can see the second term of right hand side contains scale factor which complicate our calculations, therefore in this work we consider the simplest
case with $k=0$ and $\Lambda=0$.\\
Now, we would like to solve equation (13) to obtain time-dependent energy density. Using the equation of state (5) in the energy-momentum conservation
formula (13) gives the following differential equation,
\begin{equation}\label{s15}
\dot{\rho}+3\frac{\dot{a}}{a}
\left[(1+\gamma)\rho-\frac{B-\omega-1}{(1+\omega)\rho^{\alpha}}-\frac{(\rho^{1+\alpha}-\frac{B}{1+\omega}+1)^{-\omega}}{\rho^{\alpha}}-3\zeta\frac{\dot{a}}{a}\right]=0.
\end{equation}
Then, by using the field equation (11) and change of variable $\rho=X^{2}$, one can rewrite the equation (15) approximately as the following non-linear
differential equation,
\begin{equation}\label{s16}
\dot{X}+b_{1}X^{2}+b_{2}X^{-2\alpha}+b_{3}X=0,
\end{equation}
where dot denotes derivative with respect to time, and we defined,
\begin{eqnarray}\label{s17}
b_{1}&=&\frac{\sqrt{3}}{2}(1+\gamma+\omega),\nonumber\\
b_{2}&=&-\frac{\sqrt{3}}{2}B,\nonumber\\
b_{3}&=&-\frac{3}{2}\zeta.
\end{eqnarray}
\section{Time-dependent energy density from exponential function method}
According to the exponential function method [52], the nonlinear differential equation (16) has the following solution,
\begin{equation}\label{s18}
X=\frac{c_{1}e^{-\kappa t}+c_{2}+c_{3}e^{\kappa t}}{e^{-\kappa t}+c+c_{4}e^{\kappa t}},
\end{equation}
where coefficients $c_{i}$ can be obtained by substituting the solution (18) in the equation (16). This gives the following coefficients,
\begin{equation}\label{s19}
c_{1}=\frac{1-b_{3}}{2b_{1}},
\end{equation}
\begin{equation}\label{s20}
c_{2}=\frac{(2b_{1}b-2b_{3}bb_{1}-b_{3}^{2}+1)c+(4b_{1}b+2b_{3}-2)e^{-\kappa t}}{2b_{1}(2b_{1}b+b_{3}+1)},
\end{equation}
\begin{eqnarray}\label{s21}
c_{3}&=&\frac{(4b_{1}^{2}b^{2}b_{3}+4b_{1}bb_{3}^{2}+4b_{3}b+4b_{1}^{2}b^{2}+b_{3}^{3}+b_{3}^{2}-b_{3}-1)c}{2b_{1}(2b_{1}b+b_{3}+1)^{2}}e^{-\kappa t}\nonumber\\
&+&\frac{(4b_{1}^{2}b_{3}b^{2}+4b_{1}b_{3}^{2}b+b_{3}^{3}+4b_{1}^{2}b^{2}-4b_{1}b-b_{3}^{2}-b_{3}+1)e^{-\kappa t}}{2b_{1}(2b_{1}b+b_{3}+1)^{2}}e^{-\kappa
t},
\end{eqnarray}
\begin{equation}\label{s22}
c_{4}=\frac{(1-4b_{1}bb_{3}-b_{3}^{2}-4b_{1}^{2}b^{2})c+(4b_{1}+2b_{3}-4b_{1}^{2}b^{2}-4b_{3}b_{1}b-b_{3}^{2}-1)e^{-\kappa
t}}{(2b_{1}b+b_{3}+1)^{2}}e^{-\kappa t},
\end{equation}
with,
\begin{equation}\label{s23}
b=(\frac{4b_{1}b_{2}}{b_{3}^{2}-1})^{\frac{1}{2\alpha}},
\end{equation}
also $c$ and $\kappa$ are free parameters. If $c=0$ then any dependence on time vanishes and dark energy density reduced to a constant and will be
conserved during time. So, in order to obtain time-dependent dark energy density we consider the case of $c=1$. On the other hand expanding Universe
suggest that $\kappa=1$ to have infinitesimal density at $t\rightarrow\infty$ limits. Hence, one can write,
\begin{equation}\label{s24}
\rho=\left[\frac{c_{1}e^{-t}+c_{2}+c_{3}e^{t}}{e^{-t}+1+c_{4}e^{t}}\right]^{2},
\end{equation}
In the Fig. 1 we draw typical behavior of energy density in terms of time for various models. Yellow line is corresponding to VGCG which is lower than
other cases. Extension of this model to VMCG denoted by cyan line. We see that the effect of $\gamma$ is increasing energy density. VGCCG presented by Red
line. We take values of constants $\gamma$ and $B$ from the Ref. [50]. Finally our interesting case of VMCCG denoted by green line which has bigger energy
density of other cases.\\
We find that the effect of viscosity is increasing of dark energy density. On the other hand decreasing of parameter $\alpha$ increases the value of dark
energy density. It is also found that increasing of cosmic parameter $\omega$ increases the value of the dark energy density.\\
In all cases, as expected, the dark energy density is decreasing function of time which is agree with expansion of Universe.\\
In summary, we can conclude that the evolution of the dark energy density with time is faster than other models.\\
In the Fig. 2 we give plot of the dark energy density in VMCCG model for $\alpha=0.1$ which suggests that the value of the dark energy density is very
high.

\begin{figure}[th]
\begin{center}
\includegraphics[scale=.3]{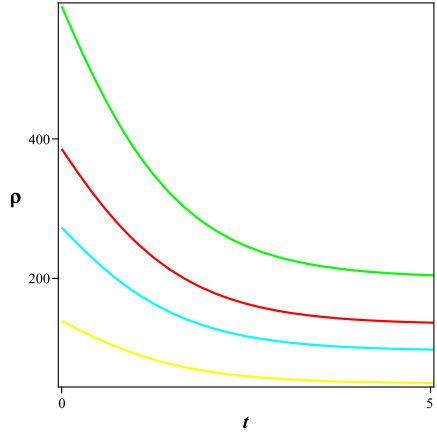}
\caption{Time-dependent dark energy density. Green: VMCCG with $\gamma=0.3$, and $B=3.4$ [50], $\omega=0.5$, $\alpha=0.5$ and $\zeta=1$. Red: VGCCG with
$\gamma=0$, and $B=3.4$ [50], $\omega=0.5$, $\alpha=0.5$ and $\zeta=1$. Cyan: VMCG with $\gamma=0.3$, and $B=3.4$ [50], $\omega=0$, $\alpha=0.5$ and
$\zeta=1$. Yellow: VGCG with $\gamma=0$, and $B=3.4$ [50], $\omega=0$, $\alpha=0.5$ and $\zeta=1$.}
\end{center}
\end{figure}

\begin{figure}[th]
\begin{center}
\includegraphics[scale=.2]{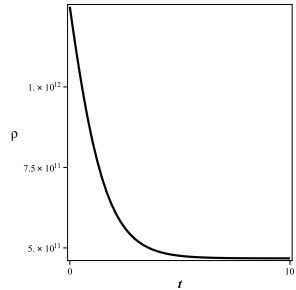}
\caption{Dark energy density of VMCCG with $\gamma=0.3$, and $B=3.4$ [50], $\omega=0.5$, $\alpha=0.1$ and $\zeta=1$.}
\end{center}
\end{figure}

\section{Hubble parameter and Scale factor}
By using the equation (24) and (11) we can find behavior of the Hubble expansion parameter. Numerically we draw Hubble parameter in terms of time in the
Fig. 3 (left), and in terms of viscous parameter $\zeta$ in the Fig. 3 (right). We can see that there is a singular point for a suitable value of viscous coefficient.\\
Time-dependent energy density (24) suggests that the scale factor takes the following general form,
\begin{equation}\label{s25}
a(t)=(me^{t}+n)^{\delta}e^{-\sigma t},
\end{equation}
where we used the relation $H=\dot{a}/a$. Equation (25) contains new constant parameters as $m$, $n$, $\delta$, and $\sigma$ which will be fixed by using
observational data. We draw typical behavior of scale factor (25) in terms of time in the Fig. 4. We can see that Hubble expansion parameter and scale
factor are decreasing function of time.\\
Before we try to fix new parameters and obtain scale factor exactly, it is useful to check stability of this theory which is subject of the next section.
\begin{figure}[th]
\begin{center}
\includegraphics[scale=.25]{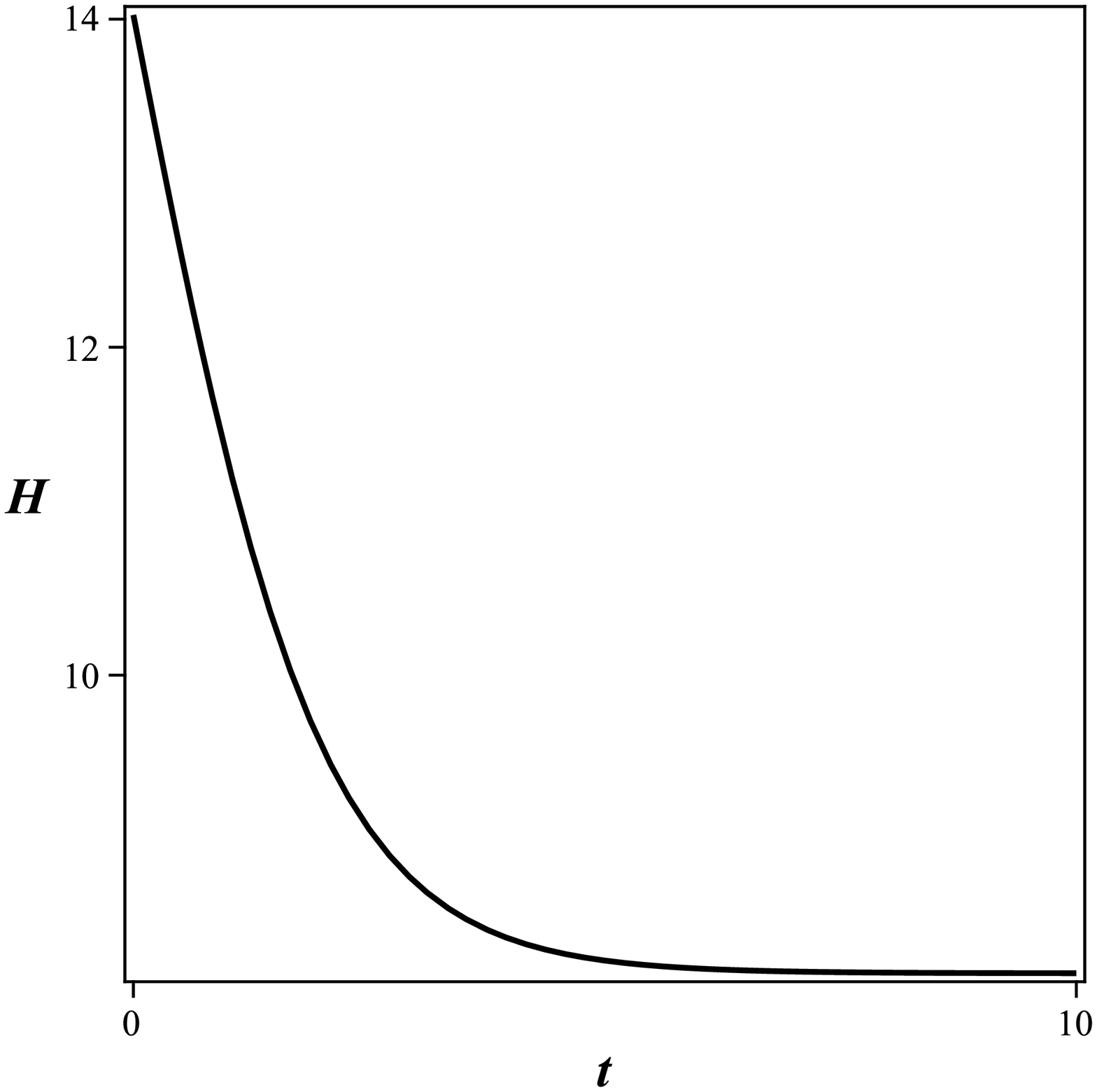}\includegraphics[scale=.25]{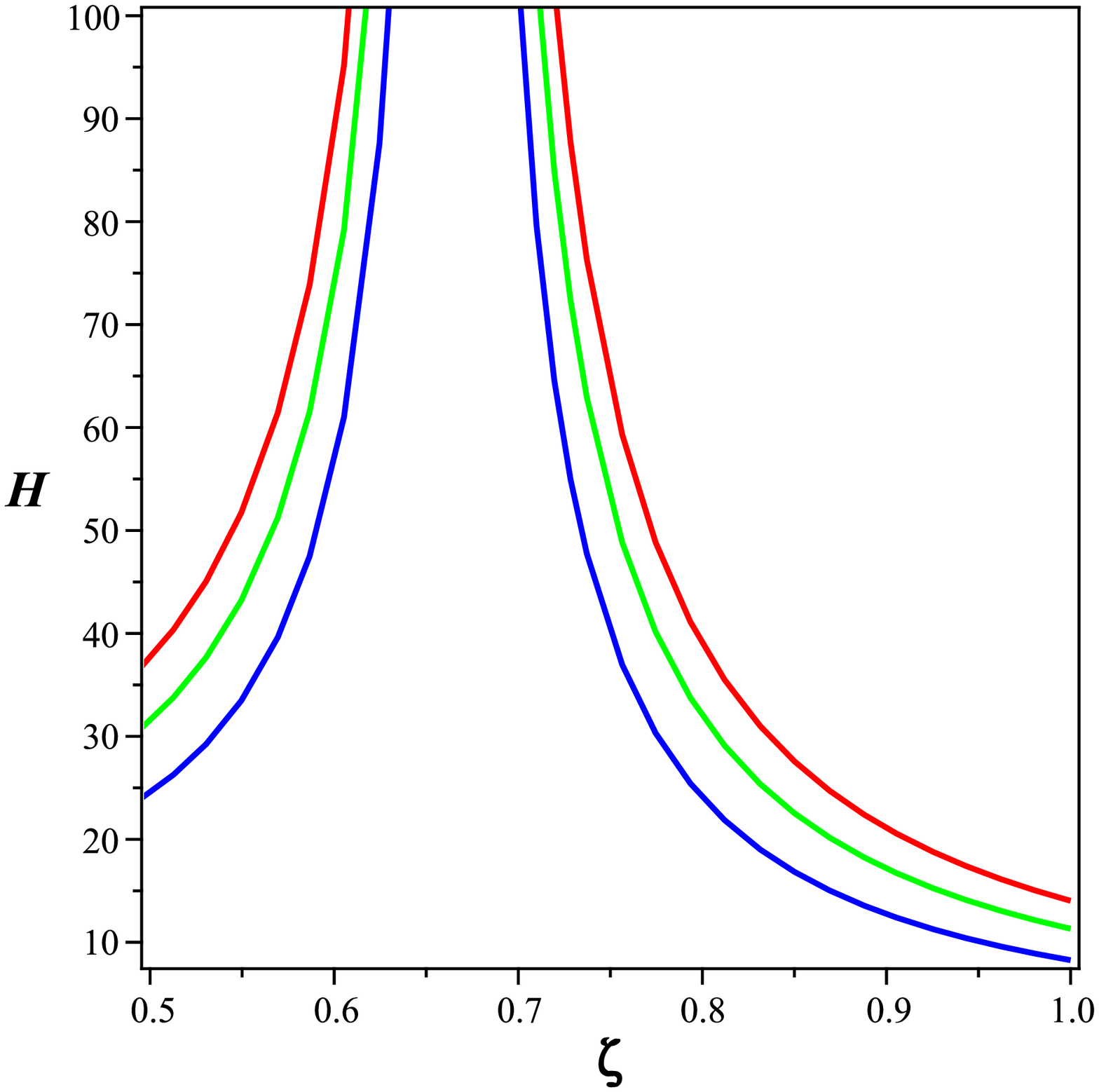}
\caption{Hubble expansion parameter in VMCCG model with $\gamma=0.3$, $B=3.4$, $\omega=0.5$ and $\alpha=0.5$. Left: In terms of time with $\zeta=1$. Right:
In terms of $\zeta$ with $t=0$ (Red line), $t=1$ (Green line), and $t=5$ (Blue line).}
\end{center}
\end{figure}

\begin{figure}[th]
\begin{center}
\includegraphics[scale=.25]{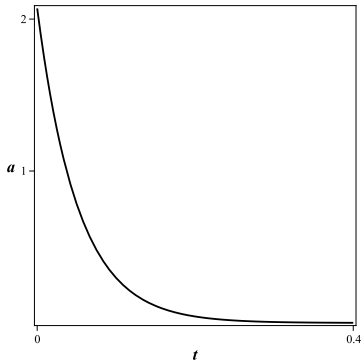}
\caption{Scale factor in VMCCG model.}
\end{center}
\end{figure}

\section{Stability}
There are several ways to investigate stability of a theory. In this paper we use reality of sound speed,
\begin{equation}\label{s26}
C_{s}^{2}=\frac{d\bar{p}}{d\rho}\geq0.
\end{equation}
Numerically, we draw $C_{s}^{2}$ in terms of time in the Fig. 5. It shows that the VMCCG model of cosmology is completely stable and there is no instable
region. In the Fig. 6, we draw $C_{s}^{2}$ in terms of $\omega$ and $\zeta$. The left plot of the Fig. 6, shows that increasing $\omega$ reduced the sound
speed.
\begin{figure}[th]
\begin{center}
\includegraphics[scale=.3]{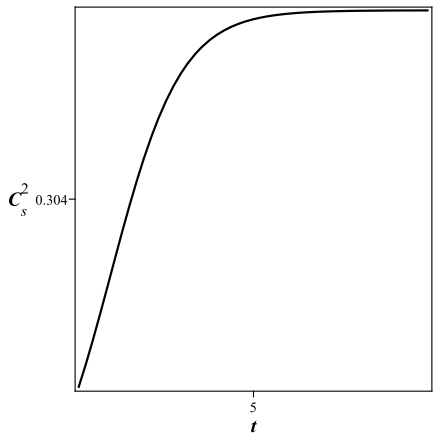}
\caption{Square of sound speed in terms of time.}
\end{center}
\end{figure}

\begin{figure}[th]
\begin{center}
\includegraphics[scale=.25]{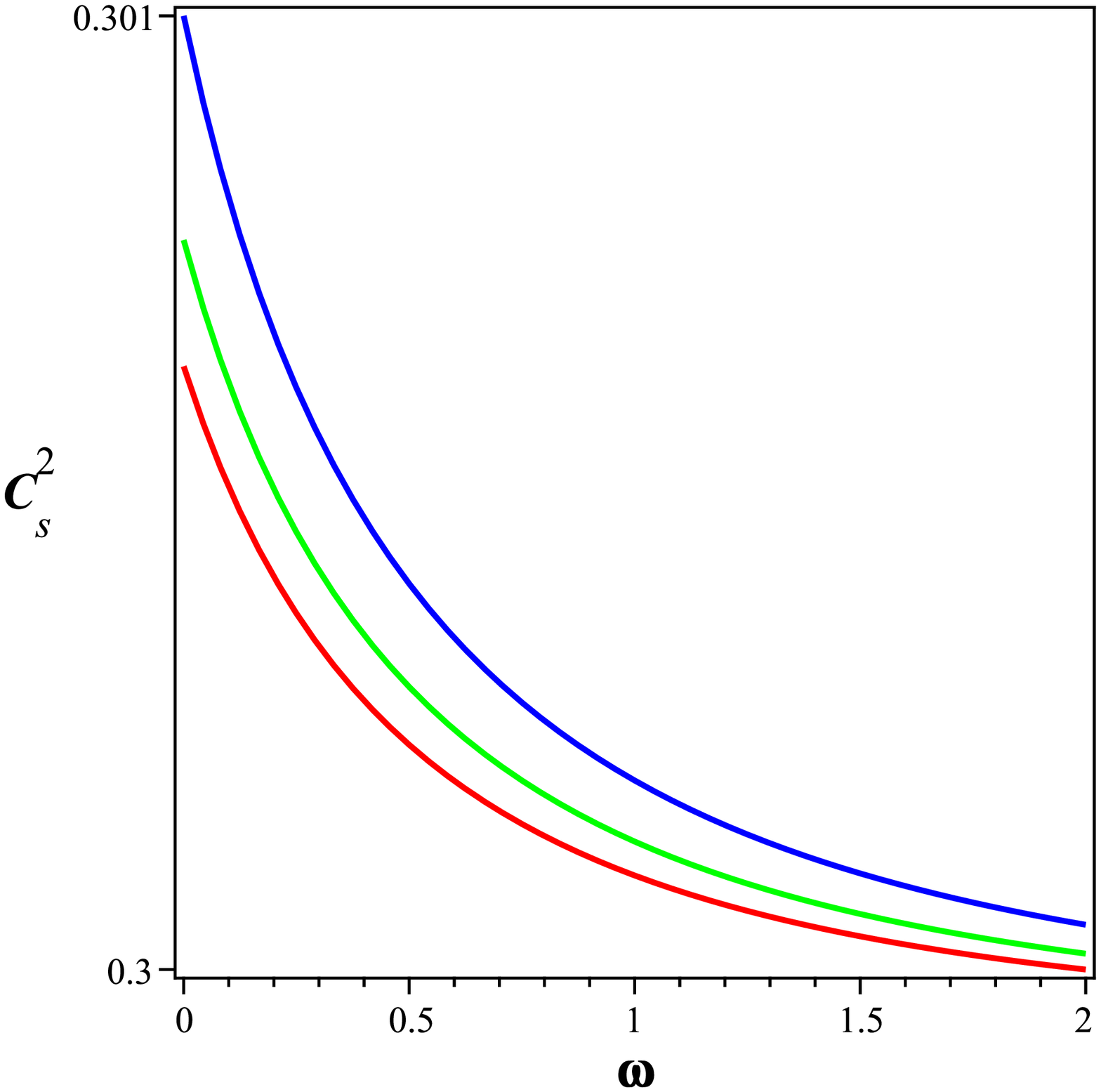}\includegraphics[scale=.25]{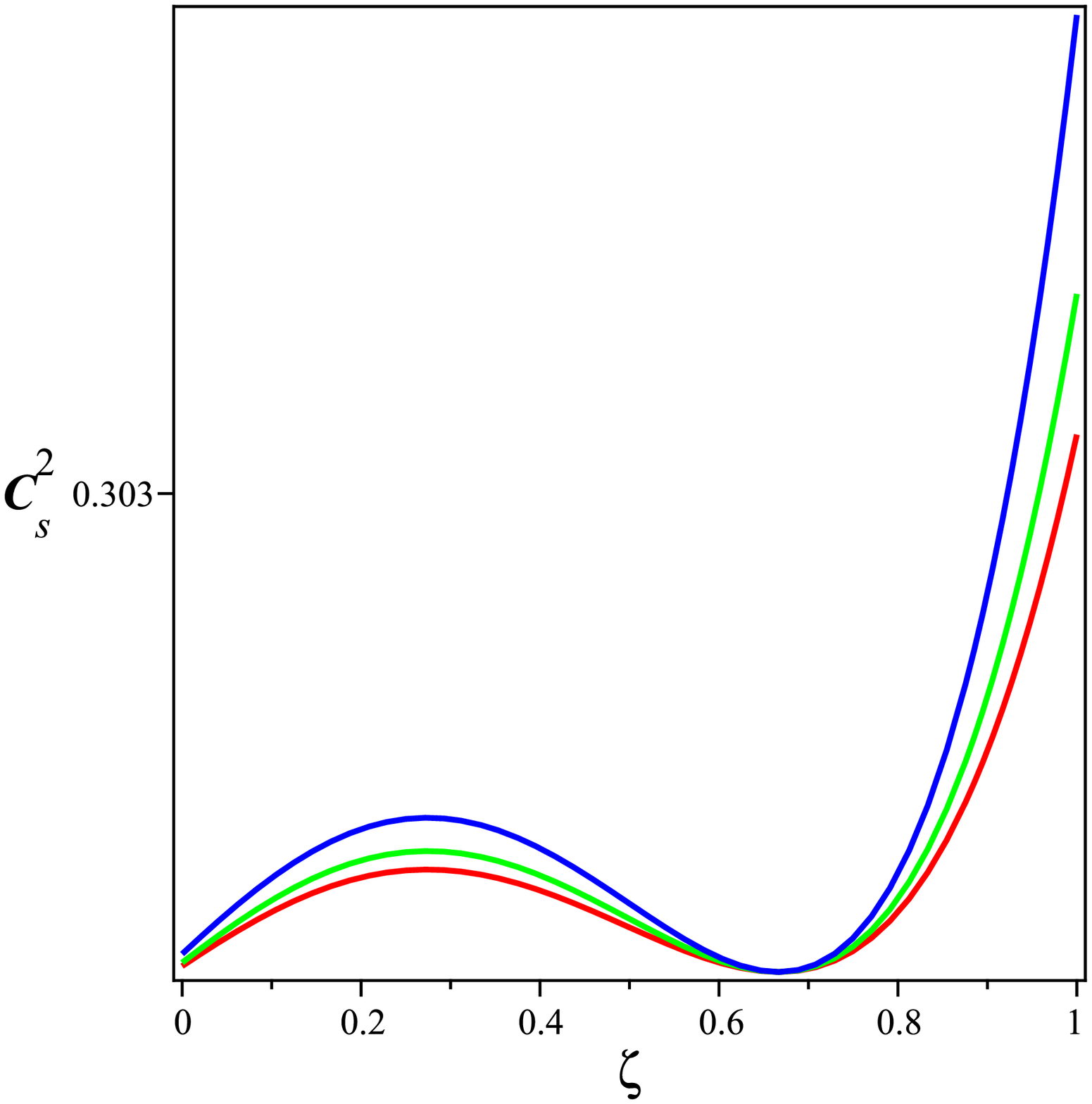}
\caption{Square of sound speed with $\gamma=0.3$, $B=3.4$, and $\alpha=0.5$. Left: In terms of $\omega$ with $\zeta=1$. Right: In terms of $\zeta$ with
$\omega=0.5$. $t=0$, $t=1$, and $t=5$ represented by Red, Green and Blue lines respectively.}
\end{center}
\end{figure}
\section{Observational data}
First of all we consider current value of Hubble expansion parameter which is $H_{0}\approx68-72$ [53]. Therefore, according to the right plot of the Fig.
3 (red line), visous coefficient $\zeta$ should be about 0.6 or 0.75, so our results agree with observational data.\\
In order to fix the scale factor (25) we use the following observational parameters [54]. Declaration parameter is given by the following relation,
\begin{equation}\label{s27}
q=-\frac{a}{\dot{a}^{2}}\frac{d^{2}a}{dt^{2}},
\end{equation}
and jerk parameter is given by,
\begin{equation}\label{s28}
j=\frac{a^{2}}{\dot{a}^{3}}\frac{d^{3}a}{dt^{3}},
\end{equation}
then, snap parameter obtained by using the following relation,
\begin{equation}\label{s29}
s=-\frac{a^{3}}{\dot{a}^{4}}\frac{d^{4}a}{dt^{4}},
\end{equation}
also, lerk parameter is given by,
\begin{equation}\label{s30}
l=\frac{a^{4}}{\dot{a}^{5}}\frac{d^{5}a}{dt^{5}}.
\end{equation}
Current value of these parameters $q_{0}$, $j_{0}$, $s_{0}$, and
$l_{0}$ appear in the following Taylor expansion around $a_{0}$,
\begin{equation}\label{s31}
\frac{a}{a_{0}}=1+H_{0}t-\frac{1}{2}q_{0}H_{0}^{2}t^{2}+\frac{1}{6}j_{0}H_{0}^{3}t^{3}-\frac{1}{24}s_{0}H_{0}^{4}t^{4}+\frac{1}{120}l_{0}H_{0}^{5}t^{5}+\cdots.
\end{equation}
In that case on can obtain,
\begin{equation}\label{s32}
q_{0}=-1-\frac{mn\delta}{(m(\delta-\sigma)-\sigma n)^{2}}.
\end{equation}
Following we use results of the Refs. [54-58] to fix our parameters.
\subsection{SNeIa}
According to the SNeIa observational data [55, 56] the current value
of the declaration parameter may be $q_{0}=-1$. This situation
obtained by three separated ways of $m=0$, $n=0$ or $\delta=0$. In
all cases one can obtain $q=s=1$ and $j=l=-1$. It means that
observational parameter have no any time dependence. It should be
noted that $q_{0}=-1$ and $j_{0}=1$ are completely agree with the
Refs. [55, 56].
\subsection{$\Lambda$CDM}
According to the $\Lambda$CDM observational data [57, 58] the
current values of the declaration and jerk parameters may be
$q_{0}=-0.6$ and $j_{0}=-1$ respectively. Scale factor (25) in the
jerk parameter (28) gives the following expression,
\begin{equation}\label{s33}
j_{0}=1-\frac{mn\delta(3n\sigma-3m(\delta-\sigma)+m-n)}{(m(\delta-\sigma)-\sigma
n)^{3}}.
\end{equation}
Assuming $\sigma=\delta$ gives us,
\begin{equation}\label{s34}
q_{0}=-1-(\frac{m}{n})\sigma^{-1},
\end{equation}
which together with $q_{0}=-0.6$ yields to,
\begin{equation}\label{s35}
\frac{m}{n}=-0.4\sigma.
\end{equation}
Combining the relation (35) with the current jerk parameter (33)
together with observational data $j_{0}=-1$ tell us that
$\sigma\approx-0.4$. Therefore on can obtain,
\begin{equation}\label{s36}
a(t)=\frac{e^{0.4 t}}{\left[n(0.16e^{t}+1)\right]^{0.4}}.
\end{equation}
With $n=0.1$ the current value of scale factor coincide with the
Fig. 4.
\subsection{Best fitted}
According to the best fitted parameters of the Ref. [54] we know
$q_{0}=-0.64$ and $j_{0}=1.02$. These observational data together
with the equations (32) and (33) give the following equation,
\begin{equation}\label{s37}
m^{2}+n^{2}+(25\delta-2)mn=0,
\end{equation}
which suggests the following solution,
\begin{equation}\label{s38}
\frac{m}{n}=\frac{25\delta-2}{2}\left[-1\pm\sqrt{1-\frac{4}{(25\delta-2)^2}}\right].
\end{equation}
We can use this relation with variation of $\delta$ to obtain more
exact solution. In the table 1, we can see that $\delta=2$ gives the
best solution relative to the best fitted data [54], and yields to
obtain the following time-dependent scale factor,
\begin{equation}\label{s39}
a(t)=a(0)(1-0.02e^{t})^{2}e^{-0.299 t},
\end{equation}
where we assumed $a(0)=n^{2}$.

\begin{table}[th]
\begin{center}
\includegraphics[scale=.75]{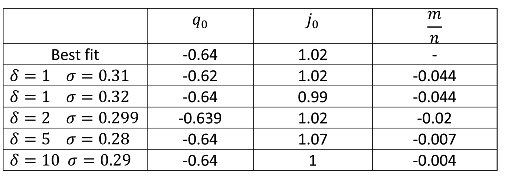}
\caption{Best fitted values of the deceleration and jerk parameters
[54], and our theoretical values. It shows that $\delta=2$, and
$\sigma=0.299$ have most agreement with observational data.}
\end{center}
\end{table}

\section{Conclusion}
In this paper we constructed viscous modified cosmic Chaplygin gas.
We reviewed evolution of Chaplygin gas to modified cosmic Chaplygoin
gas which considered as dark energy, and written corresponding
equation of state. Then, we introduced viscosity to the system and
obtained modified conservation equation and field equations. These
yield to a non-linear differential equation which gives the
time-dependent dark energy density. we solved this equation by using
exponential function method, and study behavior of the dark energy
density. Then we calculated scale factor and discussed Hubble
parameter numerically. We also investigated stability of the theory
and found that  viscous modified cosmic Chaplygin gas model is
completely stable and there is no instable region. Calculating the
scale factor yields to unknown coefficients which determined by
using the observational data of deceleration and jerk parameters. In
order to find more exact solutions one can consider also snap and
lerk parameters.\\
In this paper we considered the case of flat space ($k=0$) and
zero-cosmological constant. It is possible to include these
parameters. In that case the field equation modified as the
following,
\begin{equation}\label{s40}
H^{2}=\frac{\rho}{3}-\frac{k}{a^{2}}+\frac{\Lambda}{3}.
\end{equation}
In order to study effect of space curvature and cosmological
constant one should use the equation (40) in the relation (13) to
obtain dark energy density. The most problem is the second term of
right hand side. So, the case of $k=0$ and $\Lambda\neq0$ is
straightforward. Just enough to replace $\rho$ with
$\rho+\Lambda$.\\\\
{\bf Acknowledgments} It is pleasure to thank M. R. Ganji and A. A.
Seifi for introducing methods of solving non-linear differential
equations.

\end{document}